\documentclass[aps,prd,twocolumn,nofootinbib,superscriptaddress,longbibliography]{revtex4-2}
\usepackage{graphicx}  
\usepackage{dcolumn}   
\usepackage{bm}        
\usepackage{amssymb,amsfonts,amsmath,physics}   
\usepackage{cancel}
\usepackage{mathtools}
\usepackage{slashed}
\usepackage[dvipsnames]{xcolor}

\usepackage[bookmarks, breaklinks, colorlinks,urlcolor=blue, citecolor=red, 
linkcolor=blue]{hyperref}
\usepackage[normalem]{ulem}

\hypersetup{
	colorlinks=true,
	linkcolor=Red,
	filecolor=magenta,
	urlcolor=Blue,
	citecolor=PineGreen}

\begin{document}
\title{Probing leptogenesis and the minimal neutrino seesaw mechanism through gravitational waves}

\author{Arghyajit Datta}
\email{arghyad053@gmail.com}
\affiliation{Laboratory for Symmetry and Structure of the Universe, Department of Physics, Jeonbuk National University, Jeonju 54896, Republic of Korea}

\author{Arunansu Sil}
\email{asil@iitg.ac.in}
\affiliation{Department of Physics, Indian Institute of Technology Guwahati, Assam-781039, India}

\begin{abstract} 
We propose that a gravitational wave can be generated during leptogenesis in the early Universe which occurs during heavy right-handed neutrino decay.
Such a gravitational wave, as remnant of leptogenesis, is shown to be associated with distinguishing signatures that not only act as a powerful probe to leptogenesis but also to the existence of heavy seesaw states which otherwise remain difficult to validate despite its success in explaining the baryon asymmetry of the Universe, closely  connected to the origin of neutrino mass. 
\end{abstract}
\maketitle

The observed dominance of matter over antimatter is one of the most intriguing problems in particle physics and cosmology that cannot be explained in the realm of the standard model (SM) alone. Leptogenesis~\cite{Fukugita:1986hr,Luty:1992un,Pilaftsis:1997jf,Ma:1998dx,Hambye:2000ui,Hambye:2003ka} is perhaps the most compelling mechanism to explain such asymmetry due to its close proximity in elucidating another unsolved mystery, the neutrino mass generation. In its simplest version, the central role is generally played by the introduction of two or 
more heavy right-handed neutrinos (RHNs) $N_i$ to the SM, having the Lagrangian 
\begin{align}
	-\mathcal{L_{\rm N}}= \overline{\ell}_{L_\alpha} (Y_{\nu})_{\alpha i} \tilde{H} N_{i}+ \frac{1}{2}  \overline{N_{i}^c}M_{i} N_i + h.c.,
	\label{seesaw}
\end{align}
(in the charged lepton diagonal basis) with $\alpha = e, \mu, \tau$ and $i=1,~2.$. While their heaviness ($M \gg m_D = Y_{\nu}v/{\sqrt{2}}$, with $v = 246$ GeV being electroweak vev) is crucial to explain the smallness of light neutrino mass, $m_{\nu} = - m_D M^{-1} m_D^T$,
via type-I seesaw~\cite{Minkowski:1977sc,Yanagida:1979as,Yanagida:1979gs,GellMann:1980vs,Mohapatra:1979ia,Schechter:1980gr,Schechter:1981cv,Datta:2021elq}, the same with respect to the temperature of the thermal bath ($M \gtrsim T$) in the early Universe is instrumental for their out-of-equilibrium decay into the SM 
lepton ($\ell_{L_{\alpha}}$) and Higgs ($H$) doublets leading to the leptogenesis. For standard thermal leptogenesis, the lightest RHN (say $N_1$) responsible for generating the adequate asymmetry should satisfy the Davidson-Ibarra bound\footnote{For nonthermal leptogenesis~\cite{Lazarides:1991wu,Murayama:1992ua,Kolb:1996jt,Asaka:1999yd,Hahn-Woernle:2008tsk,Barman:2021tgt}, this bound becomes $M \gtrsim 10^6$ GeV.}: $M_1 \gtrsim 10^9$ GeV~\cite{Davidson:2002qv}. On the other hand, there prevails an upper bound  on the heaviness of the seesaw state: $M_1 \lesssim 10^{13}$ GeV above which the lepton-number-violating (by two units)
process $\ell_L + H \rightarrow \bar{\ell}_L + H^\dagger$ remains in thermal equilibrium during the decay of RHN, thereby causing a complete erasure of the asymmetry produced. 

RHN of such a high scale is inaccessible to terrestrial experiments and hence, keeps the leptogenesis and neutrino seesaw mechanism essentially away from being tested. In this letter, we find that this could actually be a blessing in disguise, as a gravitational wave (GW) can be emitted during such decay of heavy RHNs, thanks to the inevitable minimal coupling of SM sectors to gravity. In general, the 
study of GWs at ongoing and/or proposed detectors~\cite{Seto:2001qf, KAGRA:2013rdx, 2017arXiv170200786A,Reitze:2019iox,Herman:2020wao,Berlin:2021txa, Herman:2022fau,Berlin:2023grv}
provides an excellent opportunity in exploring the very early Universe~\cite{Kosowsky:1992vn,Matarrese:1993zf,Martin:1996ea,Pen:2015qta,Hindmarsh:2016lnk,Ringwald:2020ist,Domenech:2023jve,Barman:2023icn,Roshan:2024qnv,Choi:2024ilx, Datta:2025wfh} as it is essentially unaffected by the happenings during the evolution of the Universe. Here we propose that a single graviton emission can take place via {\it bremsstrahlung} during the 
decay of RHNs which can in principle reveal the characteristics of leptogenesis (occurring at a high scale) as well as the heavy seesaw states, hitherto unexplored in the literature.

\begin{figure}[t]
	\centering
	\includegraphics[width=0.35\textwidth]{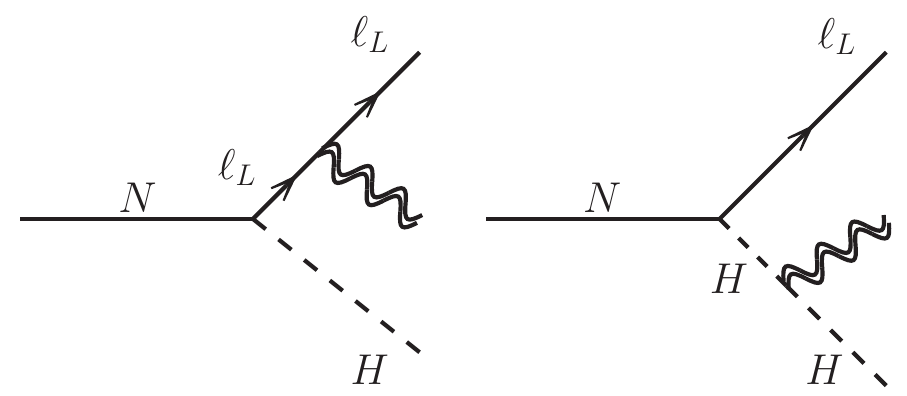}
	\caption{Feynman diagrams for GW production  via {\it bremsstrahlung}.}
	\label{fig:F-diagram}
\end{figure}

The necessary interaction terms, responsible for production of such GWs, involving the graviton, RHNs and the SM fields follow from the Einstein-Hilbert action, minimally coupled to 
gravity, of the form
\begin{align}
S = \int d^4x\sqrt{-g} \left[ 2\kappa^{-2} \mathcal{R} + \mathcal{L}_{\rm SM} + \mathcal{L_{\rm N}}\right],
\end{align}
where $\mathcal{R}$ is the Ricci scalar, with $\kappa = 2/{M_P}$ with $M_P = 2.8 \times 10^{18}$ GeV being the reduced Planck scale. Then using the weak field approximation of the metric, $g_{\mu \nu} = \eta_{\mu \nu} + \kappa h_{\mu \nu} +...$, and retaining terms of first order in $\kappa$, a coupling of canonically normalized graviton $h_{\mu \nu}$ with the stress-energy tensor $T^{\mu\nu}_X$ of SM fermion doublets/singlets and scalar (Higgs doublet here) of the form~\cite{Choi:1994ax,Holstein:2006bh}
\begin{align}
	\mathcal{L}^g_{\rm int}= -\frac{\kappa}{2} h_{\mu\nu} T^{\mu\nu}_X,  
	\label{grav-T-mu-nu}
\end{align}
results. In general, the stress-energy tensors for a fermion ($X = \psi$) and a scalar ($X = s$) are given by 
\begin{align}
	&T^{\mu\nu}_{\psi}= \frac{i}{4} \left[\bar{\psi}\gamma^\mu \partial^\nu \psi+ \bar{\psi} \gamma^\nu \partial^\nu \psi\right]- \eta^{\mu\nu} \left[\frac{i}{2}\bar{\psi}\gamma^\alpha\partial_\alpha \psi -m_{\psi} \bar{\psi} \psi\right], \nonumber \\
	&T^{\mu\nu}_{s}= \partial^\mu s\partial^\nu s-\eta^{\mu\nu}\left[\frac12 \partial ^\alpha s \partial _\alpha s -V(s)\right],
	\end{align}
respectively, where $V(s)$ corresponds to the scalar potential. 
With this minimal construction, RHNs can now have a three-body decay channel ($1 \rightarrow 3$) where a graviton is being emitted via the bremsstrahlung process, in addition to the usual two-body decay ($1 \rightarrow 2$) responsible for lepton asymmetry generation. The relevant diagrams for such $1 \rightarrow 3$ decays are shown in Fig.~\ref{fig:F-diagram} where the double curly line corresponds to the emitted graviton. The respective Feynman rules for such trilinear vertices involving left-handed lepton doublets (SM Higgs) and a graviton 
contributing to such $1 \rightarrow 3$ decay, and calculation of its differential decay width are included in the Appendix. 

Before we proceed for the evaluation of the spectrum of such GWs emitted during leptogenesis, it is pertinent to discuss the standard thermal leptogenesis scenario in the context of type-I seesaw Lagrangian presented in Eq.~\eqref{seesaw} so that its correlation with the emitted graviton energy density would become explicit.  
In the early Universe, the RHNs attain thermal equilibrium after being produced from the thermal bath via inverse decay (till
$T \gg M_i$) as well as from different scattering processes involving the SM fields. Subsequently, when the temperature drops to $T \lesssim M_i$, the out-of-equilibrium CP-violating decay of $N_i$ generates a finite amount of CP asymmetry, 
		\begin{align}
		\varepsilon_\ell^i= \frac{\Gamma(N_i\to \ell_L+H)-\Gamma(N_i\to \bar{\ell}_L+H^\dagger)}{\Gamma(N_i\to \ell_L+H)+\Gamma(N_i\to \bar{\ell}_L+H^\dagger)},
	\end{align}
	where the denominator denotes the total $1\to2$ decay width of the RHN $N_i$ and is given by (at tree level): 
	$	\Gamma_{N_i}= 
		 (Y_{\nu}^\dagger Y_{\nu})_{ii} M_i /8 \pi.
$
Note that the $1 \rightarrow 3$ decay of RHN (via Eq.~\eqref{grav-T-mu-nu}) being suppressed by the Planck scale does not effectively contribute to this decay width (and $\varepsilon_\ell^i$) and hence is excluded in the total decay width here. 

Assuming the minimal type-I seesaw scenario with two hierarchical RHNs (say, $M_1 \ll M_{2}$), the lepton asymmetry produced earlier from the decay of heavier 
$N_2$ gets diluted due to the prevailing production of the lightest RHN $N_1$ around $M_2>T>M_1$. 
 Consequently, the lightest RHN $N_1$ decay (around $T\lesssim M_1$) effectively contributes to the generation of a nonvanishing CP asymmetry and can be expressed as 
	\begin{align}
		\varepsilon_\ell\equiv \varepsilon_\ell^1= \frac{1}{8\pi (Y_\nu^\dagger Y_\nu)_{11}}  \text{Im}\left[(Y_\nu^\dagger Y_\nu)_{12}^2\right] \mathcal{F}\left(\frac{M_2^2}{M_1^2}\right),
	\end{align}
	where $\mathcal{F}(x)= \sqrt{x}\left[1+\frac{1}{1-x}+(1+x)\ln\left(\frac{x}{1+x}\right)\right]$ is the relevant loop function, generated as a result of the interference between one-loop diagram(s) and tree level $1\to 2$ decay. Here the structure of CP-violating neutrino Yukawa coupling matrix $Y_{\nu}$ can be extracted using Casas-Ibarra (CI) parametrization~\cite{Casas:2001sr}:
	\begin{align}
		Y_{\nu}=-i \frac{\sqrt{2}}{v} U D_{\sqrt{m}} {R}^{T} D_{\sqrt{M}}\,,
		\label{CI}
	\end{align}
	 where $U$ is the Pontecorvo-Maki-Nakagawa-Sakata matrix which connects the flavor basis with mass basis for light neutrinos. Here, $D_{\sqrt{m}}=  {\rm{diag}}(\sqrt{m_1},\sqrt{m_2},\sqrt{m_3})$ and $D_{\sqrt{M}}= {\rm{diag}}(\sqrt{M_1},\sqrt{M_2})$ represent the diagonal matrices containing the square root of light and heavy neutrino mass eigenvalues respectively while ${R}={R}(\theta)$ is an orthogonal matrix 
with $\theta$ being a complex angle. 
	
	To evaluate the exact amount of $B-L$ asymmetry generated from the decay of $N_1$ and its subsequent evolution with respect to time, one needs to solve the coupled Boltzmann equations (BEs) of the number densities of the $N_1$ and $B-L$ asymmetry by incorporating the decay (and inverse decay) of $N_1$, as given by
	\begin{align}
		&\frac{dn_{N_1}}{dt}+3 \mathcal{H} n_{N_1} = -(n_{N_1}-n_{N_1}^{\rm eq})\langle \Gamma_{1}^{1\to 2} \rangle-  n_{N_1}\Gamma^{1\to3},
		\label{eq:yn}\\
		&\frac{dn_{B-L}}{dt}+3 \mathcal{H} n_{B-L}= -\left[(n_{N_1}-n_{N_1}^{\rm eq})\varepsilon_\ell +n_{B-L}\frac{n_{N_1}^{\rm eq}}{n_l^{\rm eq}}\right]\langle \Gamma_{1}^{1\to 2} \rangle,\label{eq:ybl}
	\end{align}
	where $\langle \Gamma_{1}^{1\to 2}\rangle= \frac{K_1(z)}{K_2(z)} \Gamma_{N_1}$ is the thermal average of the decay rate of $N_1$ with $K_1,~K_2$ representing the Modified Bessel Functions of the 1st and 2nd kinds respectively while $z=M_1/T$ and $\mathcal{H}$ corresponds to the Hubble expansion parameter. Here $n_{N_1}^{(\rm eq)}= \frac{gT^3}{2 \pi^2} \left(\frac{M_1}{T}\right)^2 K_2 (M_1/T)$ is the (equilibrium) number density of $N_1$ with $g$ being the number of degrees of freedom. The second term on the rhs of Eq.~\eqref{eq:yn} involving $1 \rightarrow 3$ decay width of $N_1$ is kept, though insignificant for $N_1$ evolution, to indicate the depletion of $n_{N_1}$ due to graviton production from $N_1$ decay. Note that the corresponding inverse process is absent from the consideration that the gravitons have vanishing abundance compared to that of the elements of thermal bath initially.
	
\begin{figure}[t]
		\centering
		\includegraphics[width=0.45\textwidth]{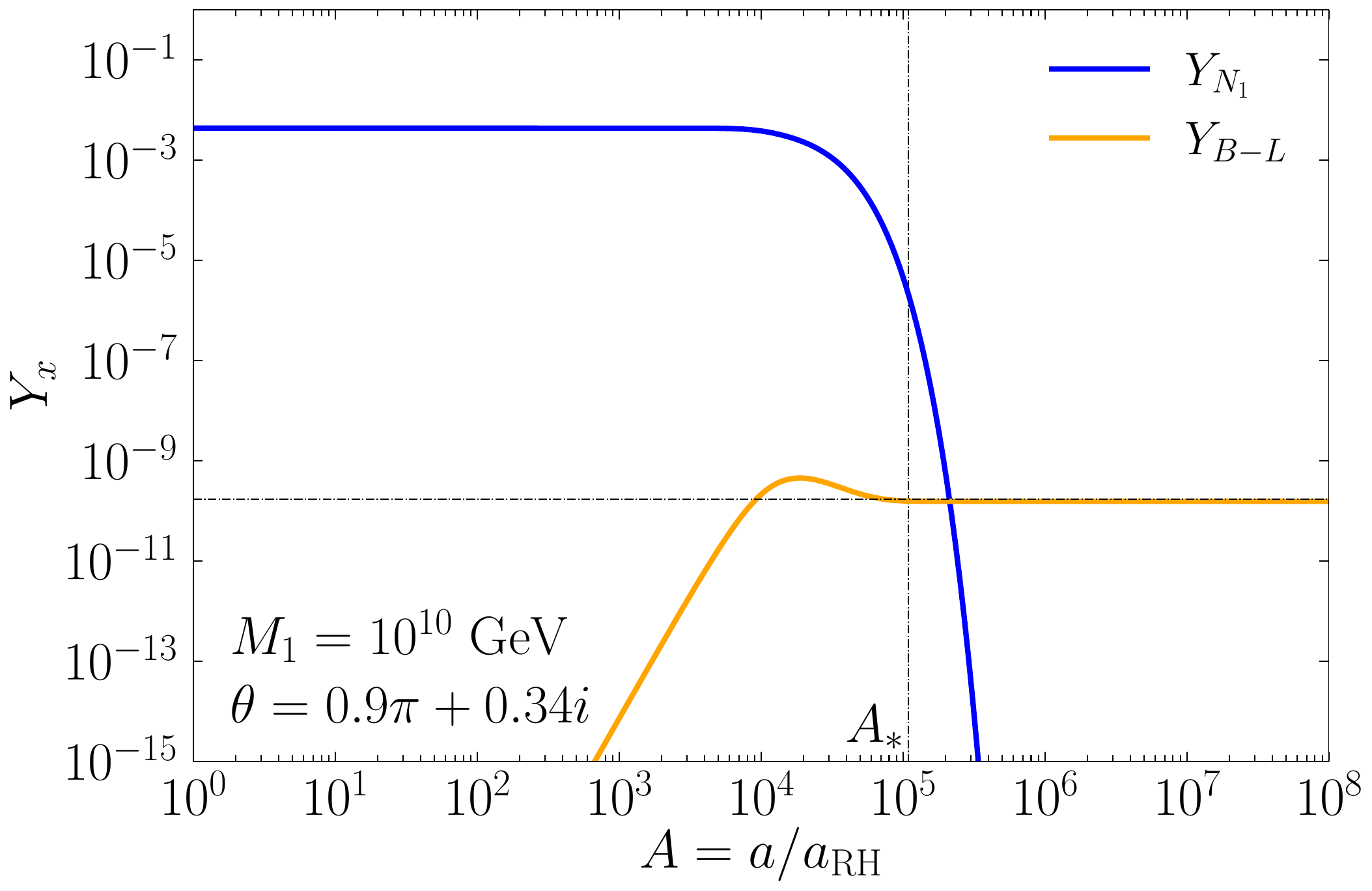}
		\caption{Evolution of $Y_{N_1}$ and $Y_{B-L}$  with respect to  scale factor $A=a/a_{\rm RH}$ for BP1. 
		\label{fig:lepto}}
	\end{figure}

For demonstration purposes, Fig.~\ref{fig:lepto} shows the variation of yields $Y_{N_1}=n_{N_1}/s,$ and $Y_{B-L}=n_{B-L}/s$  respectively ($s$ being the entropy density of the Universe), against the scale factor $A$ (normalized with respect to $a_{\rm RH}$ defined at reheating temperature $T_{\rm RH}$, assuming an instantaneous reheating\footnote{For leptogenesis during prolonged reheating, see \cite{Antusch:2006gy,Datta:2022jic, Datta:2023pav,Barman:2024jqh}.}) of the Universe\footnote{A re-parametrization of the above BEs in terms of the scalar factor $a$ can be realized via the transformation, $\frac{d}{da}\equiv\frac{1}{a \mathcal{H}} \frac{d}{dt}$.}. It corresponds to a specific choice of $M_1=10^{10}$ GeV with $\rm \theta=0.9 \pi+0.34 \it{i}$ while maintaining a hierarchy, $M_2 = 10^5 M_1$ (denoted as BP1). Note that the hierarchy among RHNs are chosen in such a way that they satisfy $M_2>T_{\rm RH}(\simeq 10^{14}~{\rm GeV})>M_1$, as a consequence of which the heavier RHNs $N_2$ are not expected to be present in thermal bath or produced during the entire evolution. The $Y_{\nu}$ is then obtained via Eq.~\eqref{CI} for normal ordered (NO) light neutrinos having $m_3>>m_2> m_1 = 0$ and using the best-fit values of neutrino oscillation parameters~\cite{Esteban:2020cvm}. In this case, the form of the $R$ matrix is given by 
 \begin{align}
 	R= \begin{pmatrix}
 		0 & \cos\theta& \sin \theta \\
 		0 & -\sin\theta & \cos \theta
 	\end{pmatrix}.
 \end{align}

\begin{figure}[t]
	\centering
	\includegraphics[width=0.45\textwidth]{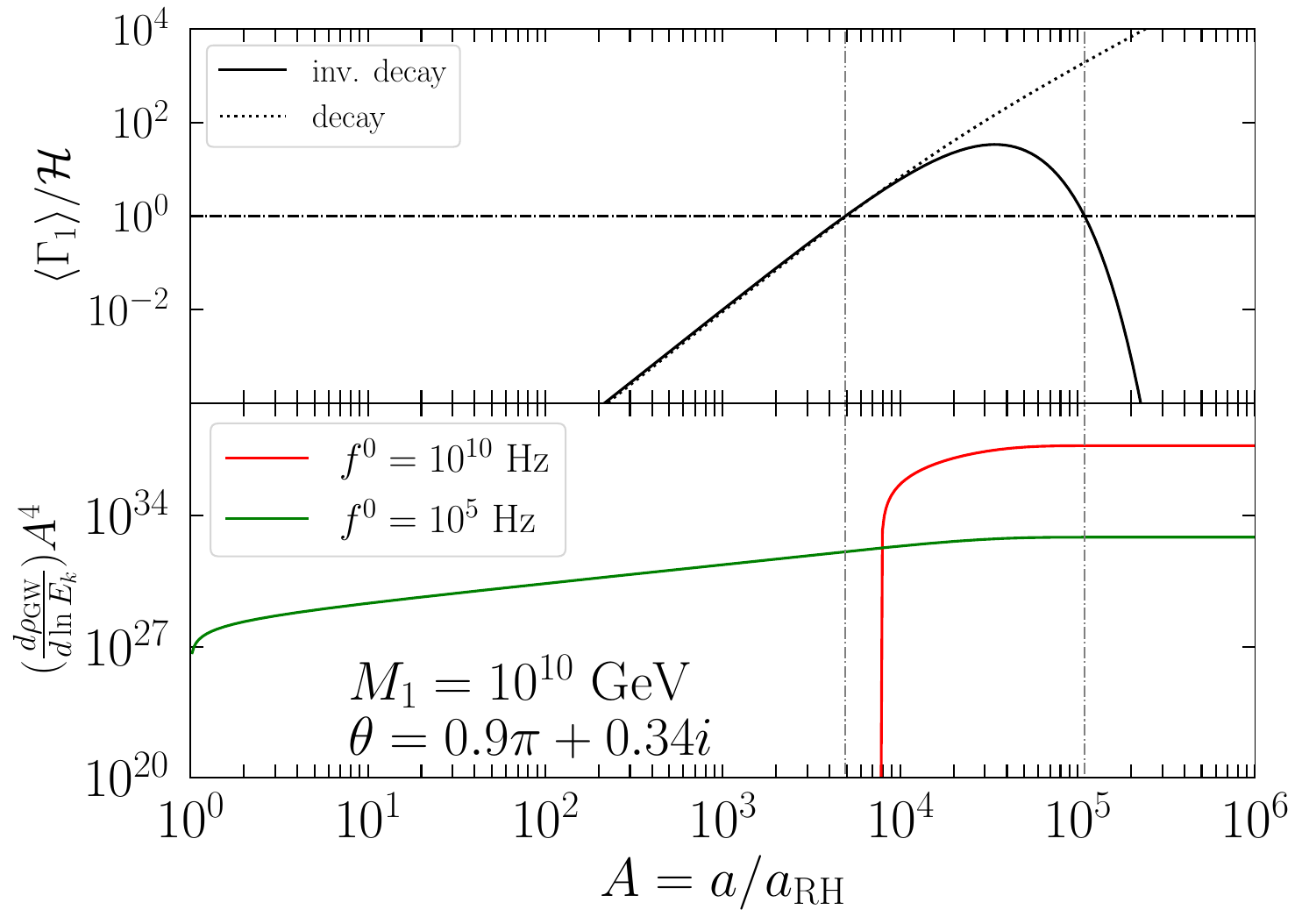}
	\caption{Evolutions of (in top panel) $\langle \Gamma_1^{\rm inv.}\rangle/\mathcal{H}$ (black solid line) and $\langle \Gamma_1^{1 \rightarrow 2}\rangle/\mathcal{H}$ (black dotted line), and (in bottom panel) comoving differential energy density of gravitons $\frac{d\rho_{\rm GW}}{d\ln E_k}A^4$ against scale factor $A=a/a_{\rm RH}$ for BP1.		
		\label{fig:lepto2}}
\end{figure}

 As stated earlier, $N_1$ remains in thermal equilibrium in the early Universe and is expected to decay predominantly at a temperature close to its mass. The actual evolution, however, is obtained by solving the BE in Eq.~\eqref{eq:yn} for the lightest RHN mass $M_1 = 10^{10}$ GeV, as illustrated by the blue curve in Fig.~\ref{fig:lepto}. As seen from the top panel of Fig.~\ref{fig:lepto2} (black solid curve), the inverse two-body decay process (denoted by  $\langle \Gamma_1^{\rm inv.}\rangle$) remains out-of-equilibrium during most of this epoch. Nevertheless, there exists a brief interval (between the two vertical dashed lines in Fig.~\ref{fig:lepto2}) during which $\langle \Gamma_1^{\rm inv.}\rangle/\mathcal{H} > 1$. As a consequence, a depletion in the $B-L$ asymmetry is observed in Fig.~\ref{fig:lepto}.
Here, the thermally averaged inverse decay rate of the RHN $N_1$ is given by
	\begin{align}
		\langle \Gamma_i^{\rm inv.}\rangle= \frac{n_{N_i}^{\rm eq}}{n_\gamma^{\rm eq}} \langle \Gamma_{N_i}\rangle	= 
		\frac{n_{N_i}^{\rm eq}}{n_\gamma^{\rm eq}} \frac{K_1(M_i/T)}{K_2(M_i/T)} \Gamma_{N_i},
	\end{align}
where $n_\gamma^{\rm eq}$ denotes the equilibrium number densities of the massless degrees of freedom. Subsequently, the two-body inverse decay process goes out-of-equilibrium, and $Y_{B-L}$ saturates around $A_* \sim 10^5$ (indicated by the vertical dashed line in Fig.~\ref{fig:lepto}), to a value corresponding to the observed baryon asymmetry of the Universe, $Y_B^{\rm exp}=8.73 \times 10^{-11}$~\cite{Planck:2018vyg}, through the relation $Y_B=\frac{28}{79}Y_{B-L}$~\cite{Harvey:1990qw}.

Note that the three-body decay of $N_1$ does not carry any direct impact on the generation and evolution of the $B-L$ asymmetry due to its origin being associated to a Planck scale-suppressed interaction [via Eq.~\eqref{grav-T-mu-nu}] compared to the sizable neutrino Yukawa coupling (responsible for two-body decay of $N_1$ producing the $B-L$ asymmetry). However, this $1 \rightarrow 3$ decay remains significant in contributing to the gravitational wave energy density produced during the $Y_{B-L}$ evolution, as we proceed to discuss below.

With the above understanding of the thermal leptogenesis scenario, we now turn our attention to obtaining the gravitational wave spectrum resulting during this leptogenesis era. To begin, we observe that the decay width of the lightest RHN toward three-body final states involving a graviton, can conveniently 
be decomposed~\cite{Barman:2023ymn, Kanemura:2023pnv} as:
	\begin{align}
		\Gamma^{1\to3} &= \int \frac{d\Gamma^{1\to3}}{d E_{k}} dE_{k}, \notag\\
		=&  \int \frac{d\Gamma^{1\to3}}{d E_{k}}  \left[\frac{M_1-E_{k}}{M_1}\right]dE_{k} + \int \frac{d\Gamma^{1\to3}}{d E_{k}}\left[\frac{E_{k}}{M_1}\right]dE_{k},
		\label{decomposed}
	\end{align}
where $E_{k}(= 2 \pi f)$ is the energy (frequency) of the graviton spanning over the range $0 < E_{k} \leq M_1/2$. The second term on the rhs
isolates the contribution of the RHN decay imparted solely to graviton, which would be helpful in determining the energy density of the GW, $\rho_{\rm GW}$. After summing over the spins and polarizations, the differential decay width for $1 \rightarrow 3$ process (see the Appendix~\ref{ap1} for the exact calculation) is found to be 
\begin{align}
		\frac{d\Gamma^{1\to3}}{d E_{k}}&= \frac{(Y_{\nu}^\dagger Y_{\nu})_{11}}{128 \pi^3} \frac{M_1^2}{M_p^2} \mathcal{G}(x),
		\label{decay-to-graviton}
\end{align}
with $\mathcal{G}(x) = (1-x)^2(1-2x)/x$ and $x = E_k/M_1$. This result is obtained in the limit of unbroken electroweak symmetry at an early Universe for which all the SM fields were massless. 
 Note that the presence of $E_k$ in the denominator of $\mathcal{G}(x)$ induces an infrared divergence in the integrated three-body decay rate of the RHNs for $E_k \rightarrow 0$. This apparent divergence, however, is systematically resolved by consistently incorporating virtual graviton loop corrections to the corresponding two-body decay process $N\to \ell_L+ H$. The interplay between real soft-graviton emission and virtual graviton corrections guarantees the cancellation of infrared divergences, thereby ensuring that all physically observable quantities remain infrared finite, as rigorously demonstrated in ~\cite{Weinberg:1965nx}.

Similar to the  lepton asymmetry calculation, here also
 the spectrum of gravitational waves is intricately related (via Eq.~\eqref{decay-to-graviton}) to neutrino Yukawa coupling $Y_{\nu}$ as well as 
the mass of the lightest seesaw state $M_1$. In other words, GW production is affected by the scale of leptogenesis due to 
its sole unavoidable out-of-equilibrium production (single graviton emission via {\it bremsstrahlung}) from $N_1$. This will be  evident as we proceed for calculation of the energy density of GW in the form of graviton radiation which satisfies the Boltzmann equation, 
\begin{align}
\frac{d \rho_{\rm GW}}{d t} +4 \mathcal{H} \rho_{\rm GW}= \left[ \int \frac{d\Gamma^{1\to3}}{d E_k}\left(\frac{E_k}{M_1}\right)dE_k\right] \rho_{N_1},
\end{align}
where $\rho_{N_1}=n_{N_1} E_{N_1}$ is the energy density of the RHN while $E_{N_1}= \sqrt{M_1^2+ 9T^2}$ represents the energy of $N_1$ in the thermal bath at $T$. Since the GW detectors are sensitive to different frequency domains, the above equation can suitably be expressed in terms of differential energy density distribution with respect to the GW energy, defined by ${d\rho_{\rm GW}}/{d \ln E_k}$, as 
	\begin{align}
		\frac{d}{d t} \left(\frac{d\rho_{\rm GW}}{d\ln E_k}\right) +4 \mathcal{H} \frac{d\rho_{\rm GW}}{d \ln E_k} =\frac
		{E_k}{M_1} \frac{d\Gamma^{1 \rightarrow 3}}{d \ln E_k} \rho_{N_1}.
		\label{eq:begw}
	\end{align}
Interestingly, the energy-weighted differential decay width, $ d\Gamma^{1\to 3}/d\ln E_k=E_k d\Gamma^{1\to 3}/dE_k$, remains finite in the soft limit, with the dependence on $E_k$ canceling out. Consequently, the differential graviton energy density, $d\rho_{\rm GW}/d\ln E_k$, is likewise free from infrared divergences. The above equation needs to be solved for $[{d\rho_{\rm GW}}/{d \ln E_k}]$  for each graviton energy $E_k$ till a point where no further GW would be generated. 
 In the present scenario, this point coincides to a stage where the Universe attained the normalized scale factor $A_*$ (at and beyond which $B-L$ asymmetry gets frozen, as stated earlier) indicative of the fact that $N_1$ decayed away completely.
Taking into account the redshifts of the energy density ($\rho_{\rm GW}$) as well as the energy ($E_k$) of the graviton, the present 
day GW energy density $\Omega_{\rm GW}^0  h^2$ can be inferred from the solution of Eq.~\eqref{eq:begw} at $A_*$, described by $[{d\rho_{\rm GW}}/{d \ln E_k}]_*$, as 
	\begin{align}
		\Omega_{\rm GW}^0 h^2 &= \left[\frac{h^2}{\rho_c^0} \frac{d \rho_{\rm GW}}{d \ln E_k}\right]_{\rm 0}=
		h^2 \left(\frac{\Omega_\gamma^0}{\rho_R^*}\right) \left[\frac{d \rho_{\rm GW}}{d \ln E_k}\right]_{ *},
		\label{eq:GW}
	\end{align}
where $\Omega_\gamma^0=\rho_R^0/\rho_c^0=5.4 \times 10^{-5}$ is the current relic density of photons and $E_{k_*}= E_k^0 (A_0/A_*)=E_k^0 (\rho_R^*/\rho_R^0)^{1/4}$ represents the energy of a single graviton at $A_*$ connected with the current energy of the same by $E_k^0 =2 \pi f^0$. 
\begin{figure}[h]
	\centering
	\includegraphics[width=0.45\textwidth]{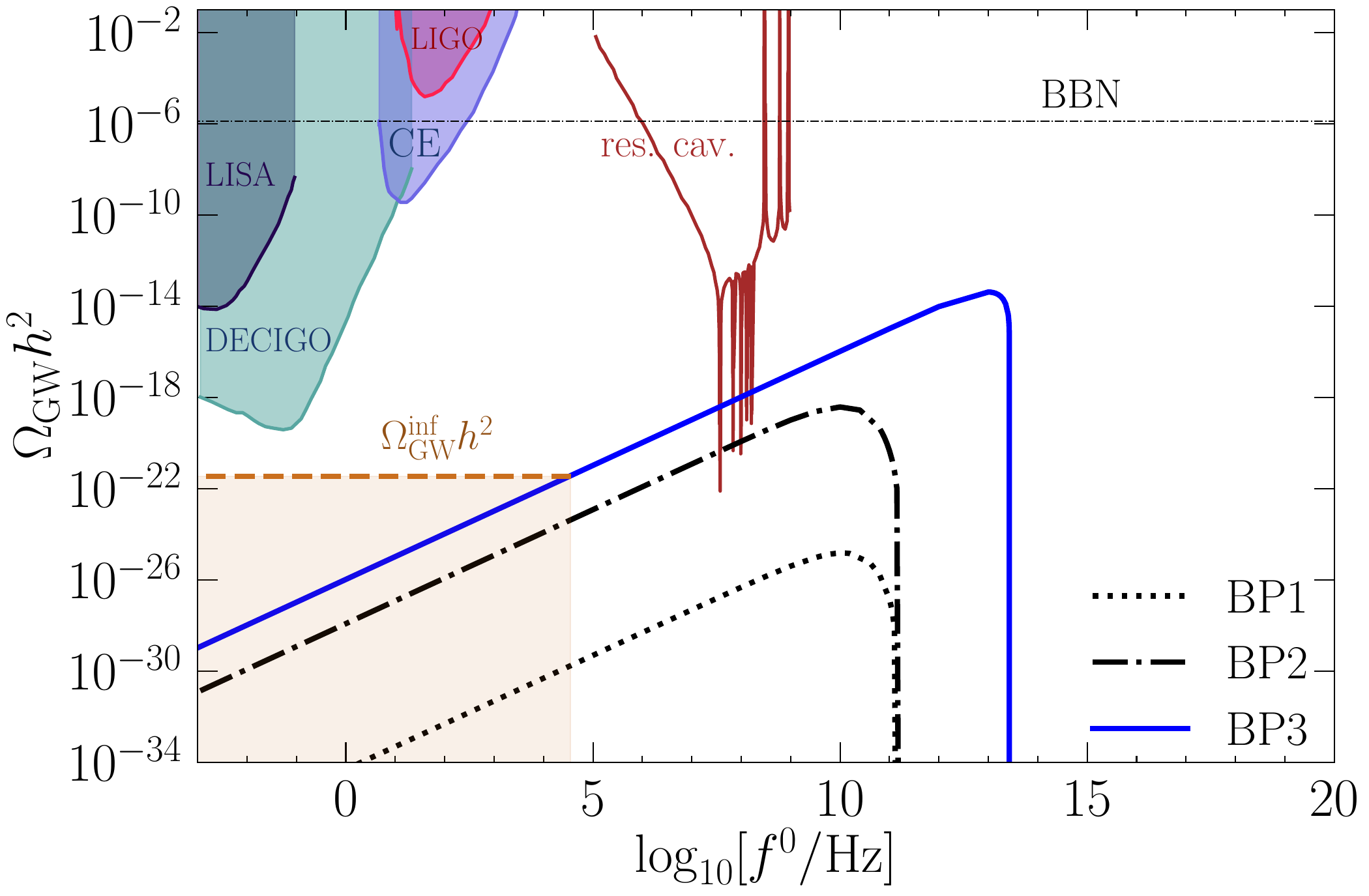}
	\caption{GW spectrum from RHN decaying to lepton doublet and Higgs when all the final state particle masses are taken to be zero. Here, BP2(3): $[M_1= 10^{13}~(10^{15})~\text{GeV},~M_2= 10^2 M_1~(5 M_1),~\theta=0.51 \pi+ 0.05{ \it i}~(0.9 \pi+ 3.8 \times 10^{-6} {\it i})]$. The bottom left pink-shaded portion indicates the region that possibly gets contaminated by the primordial GW from inflation.}
\label{fig:gw}
\end{figure}

 As shown in the bottom panel of Fig.~\ref{fig:lepto2}, gravitons with higher present-day frequencies (red line) are produced only at later cosmological times, corresponding to a larger scale factor and a lower temperature. This behavior arises because gravitons produced at earlier epochs experience a larger cosmological redshift before reaching the present epoch. Consequently, for fixed values of the present-day graviton frequency, $M_1$ and $T_{\rm RH}$, gravitons generated too early would have had energies exceeding the kinematic limit for production, namely $E_k < M_1/2$, at the time of emission. Therefore, the production of higher-frequency gravitons is restricted to later times, when the Universe has expanded further and cooled down. 

Combining with the upper panel (along with Fig.~2), we further infer that the epoch of high-frequency GW production ($f^0 \sim 10^{10} ~\mathrm{Hz}$, corresponding to the peak frequency of the GW spectrum for BP1) approximately coincides with the epoch during which inverse decay processes gradually become less efficient and eventually fall out of thermal equilibrium.  As the RHNs decay away, the graviton production process gradually becomes inefficient and effectively ceases near the completion of RHN decay. This leads to the saturation of both the comoving $B-L$ asymmetry and the comoving differential graviton energy density.

With the underlying features of graviton production in the early Universe understood, we now present our findings for the GW spectrum in Fig. \ref{fig:gw} for BP1 (BP2) where the dotted (dash-dotted) black line corresponds to the GW spectrum for $M_1 = 10^{10} (10^{13})$ GeV and NO light neutrinos with $A_* \sim 10^5~(120)$. In the same figure, we also embed future sensitivity ranges of space-based laser interferometer experiments such as LISA~\cite{2017arXiv170200786A}, DECCIGO~\cite{Seto:2001qf}, CE~\cite{Reitze:2019iox} and LIGO~\cite{KAGRA:2013rdx} working in the intermediate frequency range, spanning over $10^{-6} - 10^4$ Hz, as well as proposed resonant cavity techniques~\cite{Herman:2020wao,Herman:2022fau} possibly probing higher frequencies ranging from $10^4$ to $10^9$ Hz. We find that while the GW energy density 
$\Omega_{\rm GW}h^2$ for $M_1 = 10^{10}$ GeV falls way below the sensitivity regions of ongoing and future experiments, the one for $M_1 = 10^{13}$ GeV 
just enters into the future sensitivity region of planned resonance cavity experiment. The corresponding peak frequency is found to be $6.1 (6.7) \times 10^{10}$ Hz. Such a mild shift in peak frequency from BP1 to BP2 (while changing the mass of $M_1$ from $10^{10}$ GeV to $10^{13}$ GeV) is an artefact of the in-built changes in the neutrino Yukawa coupling $Y_{\nu}$ in order to realize correct amount of baryon asymmetry via leptogenesis,  an unexplored characteristic of GW production during leptogenesis.  A further increase in the GW energy density with $M_1 > 10^{13}$ GeV, though seems plausible by looking at the trend while moving from BP1 to BP2, is restricted in the thermal leptogenesis setup, as stated earlier. On the lower-frequency side, the IR-free behavior of the GW spectrum has been discussed already. Since our focus here is on the phenomenologically relevant frequency range, for clarity of presentation, we introduce a representative low-frequency cutoff at $f^0_{\rm min}=10^{-3}\,\mathrm{Hz}$ in Fig.~\ref{fig:gw} and do not display the deep-IR portion of the spectrum.


Notice that, in the infrared regime, the GW spectrum exhibits a linear dependence on frequency, $\Omega_{\rm GW} h^2 \propto f^0$. This behavior can be understood from the structure of the Boltzmann equation Eq.~\eqref{eq:begw} governing the evolution of the graviton energy density. In particular, as stated earlier, the $E_k$ dependence is effectively cancelled in the energy-weighted differential decay width $d\Gamma^{1\rightarrow 3}/d\ln E_k$ in the soft limit. Consequently, the additional factor of the graviton energy, $E_k \propto f$ entering on the right-hand side of the BE leads to the characteristic infrared scaling $\Omega_{\rm GW}h^2\propto f^0$.

Based on the finding above, we notice that other leptogenesis scenarios that work with lighter RHNs such as resonant leptogenesis~\cite{Pilaftsis:1997jf,Pilaftsis:2003gt}, would only produce less GW energy density and hence the GW spectrum should fall below the sensitivity region of the planned and ongoing experiments. On the other hand, for a nonthermal leptogenesis, the GWs produced via bremsstrahlung~\cite{Nakayama:2018ptw,Ghoshal:2022kqp,Barman:2023ymn,Bernal:2023wus,Choi:2024acs,Xu:2024fjl} during the decay of the heavy particle (e.g., inflaton) to RHNs~\cite{Ghoshal:2022kqp} 
would be stronger, though do not carry the characteristic signature of leptogenesis, than those generated during the subsequent decay of the RHNs during nonthermal leptogenesis. Similarly, some alternate leptogenesis scenarios where GWs are generated
due to the formation of domain walls \cite{Barman:2022yos} and cosmic strings \cite{Dror:2019syi,Chianese:2024gee} bear the features of those exotic happenings rather than carrying signatures specific to the process of leptogenesis from RHN decay.

However a situation may prevail, where the (large) masses of the RHNs find their origin associated with a phase transition (PT) in the early Universe at a temperature $T_*$. 
For example, there could be a bubble collision in case the PT being of first order, produces suddenly heavy 
RHNs (as they enter inside the bubble of true vacuum) \cite{Watkins:1991zt,Konstandin:2011ds,Falkowski:2012fb,Baldes:2021vyz, Dasgupta:2022isg,Cataldi:2024pgt} or 
there might be an interaction involving RHNs and a SM singlet scalar field $\phi$ of the form $\lambda_i \phi N_iN_i$, respecting a global $U(1)$ symmetry, for which RHNs become massive during a second-order PT at $T_*$ with a nonzero vacuum expectation value of $\phi$. In either case, provided masses of the RHNs are larger compared to $T_*$, they decay immediately. Such an instantaneous decay contributes not only to the production of lepton asymmetry but also to the production of GW via bremsstrahlung, similar to the preceding discussion. 

To proceed with such a sudden gain of mass for the RHNs related to PT, we first note that the RHNs (two here) were massless and part of the thermal bath prior to the PT, and suddenly both become massive at $T_*$. To be specific, if we consider the latter scenario described above with $\lambda_i \phi N_iN_i$ interaction, we can employ the same Eq.~\eqref{eq:begw} for finding out 
$\Omega_{\rm GW} h^2$ contributed by both $N_{1,2}$ while replacing the initial (at the onset of PT) number density of the RHNs by their relativistic equilibrium number density, $n_{N_i}^{eq}$. We keep masses of $N_{1,2}$ close enough here, so that dilution of $B-L$ asymmetry via entropy injection due to the decay of heavier $N_2$ can be restricted
We observe that $N_{1}$ of mass $10^{15}$ GeV, $M_2 = 5 M_1$ with $T_* = 10^{12}$ GeV can bring the GW spectrum well within the sensitivity range of proposed resonance cavity experiment (as shown in Fig. \ref{fig:gw} by blue solid line) while generating the observed baryon asymmetry simultaneously. Note that as $T_*$ remains below $10^{13}$ GeV, the $\Delta L =2$ process $\ell_L + H \rightarrow \bar{\ell}_L + H^\dagger$ is not in equilibrium (which prevented us to go beyond $M_1 > 10^{13}$ GeV in case of thermal leptogenesis) and hence, a complete erasure of asymmetry by such process is no longer applicable. On the other hand, it is also observed that a significant increase of $M_{1,2}$ beyond $10^{16}$ GeV would introduce sizable elements, beyond the limit of perturbativity, of the neutrino Yukawa coupling and hence not considered. 
Notice that such a phase transition may involve bubble nucleation and/or the formation of topological defects such as cosmic strings, both of which can act as GW sources~\cite{Weir:2017wfa,Caprini:2018mtu,Bae:2025gpc}. While a detailed treatment of the associated dynamics lies beyond the scope of the present work and remains an active area of research, existing phenomenological fits to numerical simulations indicate that the resulting GW spectra typically peak at significantly lower frequencies than those associated with the graviton bremsstrahlung mechanism considered here. Consequently, these additional sources are not expected to affect the distinctive high-frequency features of our predicted GW spectrum, related to leptogenesis.

  Regardless of these variations of leptogenesis setup, 
	it is perhaps important to investigate the primordial gravitational wave production arising from the quantum fluctuations that occurred during inflation. 
	Throughout our analysis we have considered the maximum temperature that the Universe can attain is $10^{14}$ GeV, i.e., $T_{\rm RH}=10^{14}$ GeV, assuming an instantaneous reheating process. This helps us evaluating the Hubble expansion rate at the end of inflation, as given by~\cite{Maggiore:1999vm, Guzzetti:2016mkm, Fujita:2020rdx}
	\begin{align}
		&{\mathcal{H}}_{\rm end} \simeq 1.43 \times 10^{10}~{\rm GeV} \left(\frac{g_*^{\rm RH}}{106.7}\right)^{1/2} \left(\frac{T_{\rm RH}}{10^{14}~{\rm GeV}}\right)^2.
	\end{align}
Identifying this Hubble rate with that of the one during the inflationary phase as it remains essentially constant during this period, one finds a scale-invariant spectrum of GW from quantum fluctuations expressed by
	\begin{align}
		&\Omega_{\rm GW}^{\rm inf}h^2 \simeq 3.5 \times 10^{-22}  \left(\frac{g_*^{\rm RH}}{106.7}\right)^{-1/2} \left(\frac{{\mathcal{H}}_{\rm end}}{1.43 \times 10^{10}~{\rm GeV}}\right)^2\\
		&{\rm with,}\notag\\
		&f^0 \leq f^0_{RH}= 4 \times 10^{4}~ {\rm Hz} \left(\frac{g_*^{\rm RH}}{106.7}\right)^{-\frac{1}{12}} \left(\frac{H_{\rm end}}{1.43 \times 10^{10}~{\rm GeV}}\right).
	\end{align}
	Consequently, we find that these gravitational waves  (as indicated by the shaded region bounded by the orange dashed line in Fig.~\ref{fig:gw}) contaminate the gravitational wave spectrum produced by the bremsstrahlung process at the lower-frequency range $f^0\leq 4 \times 10^{4}~ {\rm Hz}$ with amplitude $\Omega_{\rm GW}^{\rm inf}h^2  \simeq 3.5 \times 10^{-22}$.

Finally, to conclude, our study indicates that it is indeed possible to probe leptogenesis through GWs that were emitted in the form of graviton radiation during the decay of heavy right-handed neutrinos. In fact, the higher-frequency side (near peak frequency) of the gravitational wave spectrum arises during the leptogenesis epoch, when inverse decays of RHNs become inefficient and the RHN abundance begins to deplete, leading to the simultaneous saturation of both the $B-L$ asymmetry and the gravitational wave production rate.  
	  
We also briefly comment on the inverted ordering (IO) case in the minimal seesaw framework and compare it with the NO scenario adopted in the main text. In the IO case, the light neutrino spectrum is quasidegenerate, which can enhance the relevant Yukawa combination $(Y_\nu^\dagger Y_\nu)_{11}$ within the CI parametrization. As a result, $(Y_\nu^\dagger Y_\nu)_{11}$ being moderately larger than in the NO case for suitable choices of the complex angle $\theta$, it may lead to an enhanced gravitational wave signal. However, this enhancement is parameter dependent and remains numerically modest for the benchmark points considered. In particular, BP1 exhibits a mild increase in amplitude, BP2 remains essentially unchanged, while BP3 is constrained by perturbativity bounds on the Yukawa couplings, rendering this benchmark incompatible with the inverted ordering. 

Going beyond the current minimal setup by introducing three RHNs, the canonical  seesaw, would lead to presence of additional free parameters through the orthogonal matrix $R$ in Eq.~\eqref{CI}. Specifically, it then would contain four additional real parameters. These extra parameters enlarge the allowed Yukawa structures and are therefore expected to broaden the leptogenesis parameter space, in general.  Nevertheless, the main outcome related to the GW spectrum during leptogenesis in our analysis is expected to remain qualitatively unchanged with the strongly hierarchical RHNs assuming the masses of the two heavier RHNs remain above the maximum temperature of the thermal bath (similar to BP1 and BP2 of this setup). Since heavier RHNs cannot be produced, they do not directly affect the GW signal associated with RHN dynamics. In this case, the generation of lepton asymmetry is also essentially governed by the decay of the lightest RHN. 
Interestingly, the situation can be different in the canonical seesaw with hierarchical RHNs where RHN masses are generated dynamically via the nonzero vacuum expectation value of a scalar field during a PT (similar to the setup representing BP3). As discussed earlier, this setup effectively realizes a nonthermal leptogenesis scenario. In this case, the heavier RHNs may contribute both to leptogenesis and to the generation of additional GWs through their decays, and thereby are expected to affect the GW spectrum we predict at present.

At present, based on the proposed sensitivity range,  it turns out that the resonant cavity experiment \cite{Herman:2020wao,Herman:2022fau} is capable of detecting such gravitational waves in case the seesaw state(s) is very heavy\footnote{Although, recent analyses ~\cite{Aggarwal:2020umq} suggest that a more careful comparison between the signal and background noise can significantly modify the effective sensitivity region.}. While resonant cavities are primarily optimized for detecting sharp spectral features within their resonance bandwidth, they can also be sensitive to smooth stochastic GW backgrounds as in our work, under appropriate assumptions. Since the spectrum (as seen from Fig.~\ref{fig:gw}) varies slowly with frequency, it is effectively constant over the narrow resonance bandwidth, allowing the detector to measure the local amplitude of the signal.  Even then sensitivity can be improved through long integration times, which enhances the signal-to-noise ratio, and through networks of detectors operating at different resonance frequencies. In particular, cross-correlation between independent detectors can suppress uncorrelated noise while retaining the correlated stochastic signal, thereby improving detectability. In that case, with enhanced sensitivity range and planning of GW detectors at higher frequency range,
such probes of leptogenesis (and seesaw mechanisms) can be extended for lighter seesaw states as well.  Interestingly, the mechanism is not limited to the decay of RHNs only, rather the same can be extended to other leptogenesis scenarios~\cite{Ma:1998dx,Hambye:2000ui,Albright:2003xb,Chen:2009vx,Khalil:2012nd,AristizabalSierra:2014nzr,Rink:2020uvt,Datta:2021gyi,Vatsyayan:2022rth,Pramanick:2024gvu} involving heavy seesaw states like triplet scalars or fermions in the context of type-II~\cite{Schechter:1980gr,Lazarides:1980nt,Mohapatra:1980yp,Wetterich:1981bx} or III~\cite{Foot:1988aq} seesaw scenarios. Furthermore, as shown in a recent work by
us \cite{Bhandari:2023wit}, leptogenesis with RHNs having mass below the electroweak scale is also a possibility with temperature-dependent heavy mass of RHNs at the early Universe. Our present proposal is equally applicable to such scenarios also. Overall, the study of such GW spectrum associated to leptogenesis will open up several unexplored avenues for research in the field of leptogenesis which remains difficult to study at collider experiments because of the involvement of heavy seesaw states. Another salient feature of our work is the simultaneous probe of the heavy seesaw states involved in neutrino seesaw, more importantly as the intensity of the emitted GW crucially depends on the neutrino Yukawa coupling and the mass of the heavy seesaw states indicating an indirect probe of seesaw mechanism too. 

\begin{acknowledgements}
	The authors would like to thank Jan Sch{\"u}tte-Engel for valuable discussion over email and Rajat Kumar Mandal for stimulating conversations.
	 The work of AD is supported by the National Research Foundation of Korea (NRF) grant funded by the Korean government (MSIT) (No. NRF-2022R1A4A5030362). AD also acknowledges the support provided by the Department of Physics, Kyungpook National University during his stay at Daegu, South Korea. The work of AS is supported by the grants CRG/2021/005080 and MTR/2021/000774 from SERB, Govt. of India. 
\end{acknowledgements}
\newpage
\newpage
\onecolumngrid

\section{appendix}
\label{ap1}
\begin{figure}[b]
	\centering
	\includegraphics[width=0.4\textwidth]{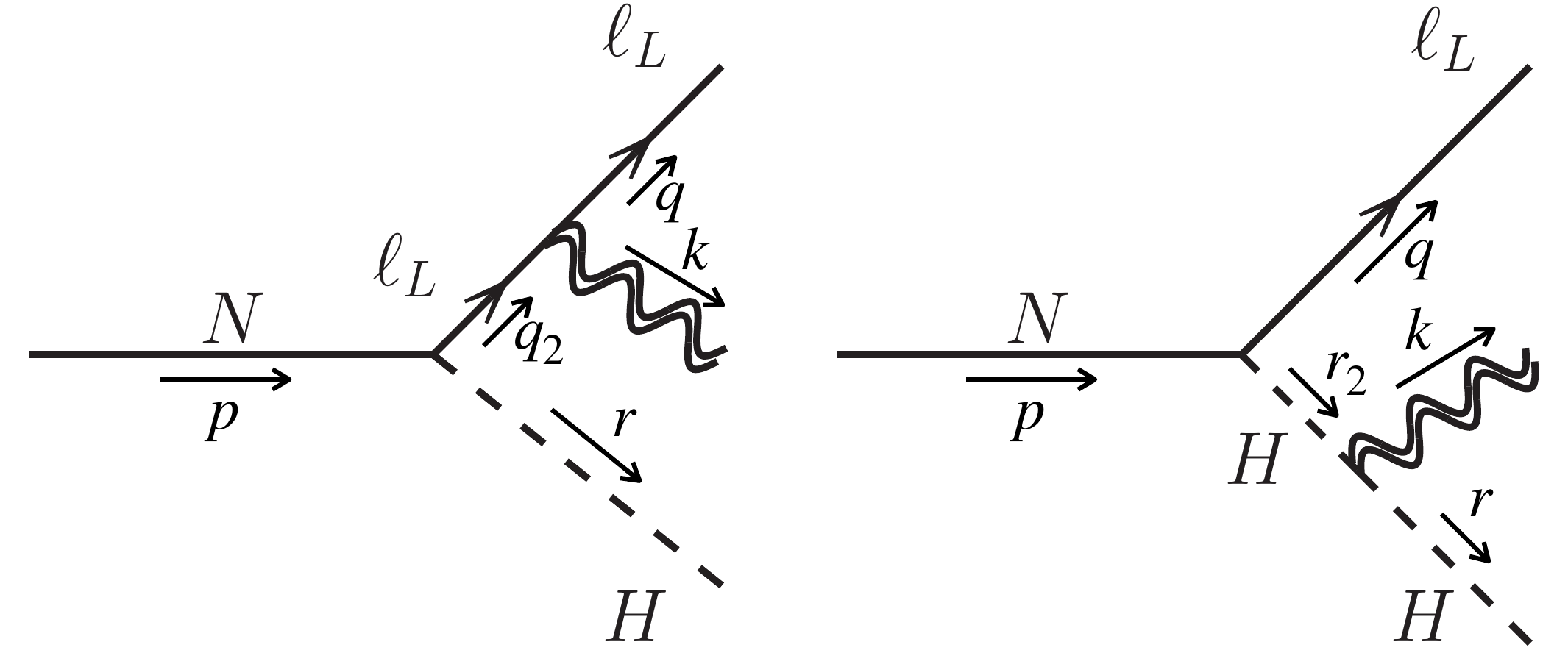}
	\caption{Feynman diagrams relevant for GW production from lepton and Higgs leg.}
	\label{fig:diagram2}
\end{figure}

Here, we plan to evaluate the differential decay rate (${d\Gamma^{1 \to 3}}/{dE_k}$) of the three-body decay process of the RHN to the lepton and Higgs doublet with the possible emission of single graviton (double curly lines) as shown 
in Fig.~\ref{fig:diagram2}.
The graviton being a massless spin-2 particle has the associated polarization tensors 
$\epsilon^{\mu\nu}_{i=1,2}$ satisfying the symmetric and transverse relations:
\begin{align}
	\epsilon_i^{ \mu \nu}&= \epsilon_i^{\nu\mu}, ~~k_\mu \epsilon_i^{\mu \nu}= 0,
	\label{eq:ep1}
\end{align}
where $k= (E_k,\vec{k})$ represents the graviton four-momentum with $k^2=0$. Furthermore, they are traceless and  orthonormal as specified by, 
\begin{align}
	\eta_{\mu\nu} \epsilon_i^{\mu\nu}&=0, \epsilon_i^{\mu\nu}\epsilon_{j_{\mu\nu}}=\delta_{ij},
	\label{eq:ep2}
\end{align}
where $\eta_{\mu\nu}$ is the flat metric. Additionally, summing  over polarization indices provides
\begin{align}
	\sum_{pol.} \epsilon^{* \mu\nu}\epsilon^{\alpha\beta}= \frac12 \left[\hat{\eta}^{\mu\alpha}\hat{\eta}^{\nu\beta} +\hat{\eta}^{\mu\beta}\hat{\eta}^{\nu\alpha}-\hat{\eta}^{\mu\nu}\hat{\eta}^{\alpha\beta} \right],~\text{with}
	\hspace{1em}
	\hat{\eta}^{\mu\nu}= \eta^{\mu\nu}- \frac{k_\mu\bar{k}_\nu+k_\nu\bar{k}_\mu}{k.\bar{k}}.
\end{align}
The massless nature of the graviton implies $k.\bar{k}=2 E_k^2$ and $\bar{k}=(E_k,-\vec{k})$.

To proceed with the evaluation of the differential decay rate, for simplification, a coordinate system is chosen in 
which the produced gravitons have momentum along $x$ direction, leading to $k=(E_k,k_x,0,0)$. Then, the 
four-momentum of the decaying RHNs can be expressed as $p= (M_i,0,0,0)$, while the four-momentum associated to $\ell_L$ and $H$ takes the form $q=(E_q,q_x,q_y,q_z),~r=(M_i-E_q-E_k,-q_x-k_x,-q_y,-q_z)$ respectively. 
With these four vectors, the following relations are obtained:
\begin{align}
	&p.p =M_i^2,~~q.q=m_l^2,~~r.r= m_H^2,\label{eq:kin1}\\
	&p.q= M_i  E_q,~~ p.r= M_i (M_i-E_k-E_q),~~p.k=p.\bar{k}= M_i E_k,\\
	&q.r=\frac12(M_i^2-2 M_i E_k-(m_l^2+m_H^2)),~~q.k= M_i(E_k+E_q-\frac{M_i}{2})+\frac{1}{2}(m_H^2-m_l^2),~~ q.\bar{k}= 2 E_q E_k- q.k,\\
	&r.k=M_i E_k - q.k,~~ q.\bar{k}=M_i E_k- 2 E_q E_k-2 E_k^2+ q.k\,,\label{eq:kin2}
\end{align}
which will be useful in calculating the differential decay width.
\begin{figure}[t]
	\centering
	\includegraphics[width=0.8\textwidth]{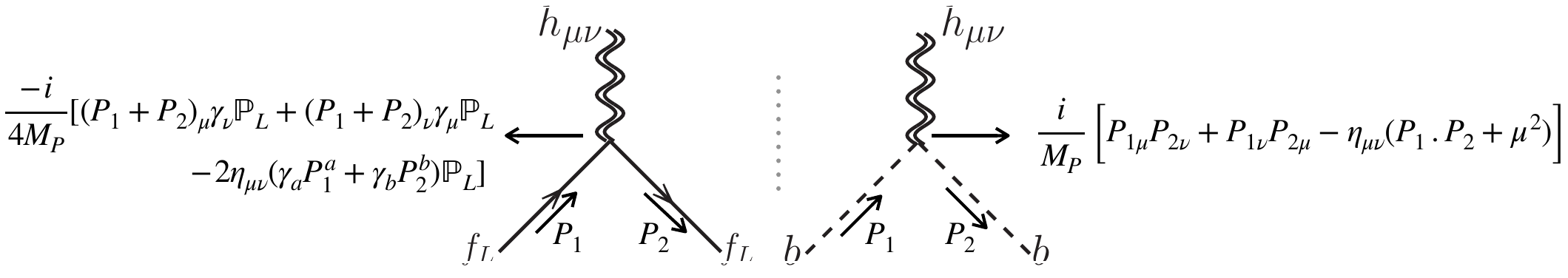}
	\caption{Feynman rules relevant for GW production from lepton and Higgs leg.}
	\label{fig:frule}
\end{figure}
We now move on to evaluate the Feynman amplitudes for both the diagrams of Fig.~\ref{fig:diagram2}. 

Since the RHNs interact only with the left-handed lepton doublets $\ell_L$ and the SM Higgs $H$ via neutrino Yukawa interaction, gravitons can only emit (in the lowest order in $\kappa =2/M_P$) from either the left-handed lepton side or the Higgs side as shown in left and right panels of Fig.~\ref{fig:diagram2} respectively. The relevant vertex factors can 
be derived from Eqs.~(3)-(4) of the main text and are presented in Fig.~\ref{fig:frule}. Using the vertex factor involving   Fig.~\ref{fig:frule} and the properties of the polarization tensor from Eq.~\eqref{eq:ep1} and \eqref{eq:ep2}, the Feynman amplitude for the $1 \rightarrow 3$ decay of RHN (graviton being emitted from the lepton side) can be estimated as
\begin{align}
&\mathcal{M}_1= -\frac{Y_{\nu} q_\mu}{2 M_p  (q.k)}\left[\bar{u}_{\ell}(q) \gamma_\nu \mathbb{P}_L (\slashed{q_2}+m_\ell)\mathbb{P}_R u_N(p)\right]\epsilon^{* \mu\nu}.
\end{align}
Here the Feynman rule for the $N-\ell_L H$ vertex is considered as $-i (Y_\nu) \mathbb{P}_R$. Moreover the  incoming Majorana field is denoted by $u_N$ following the convention of left to right fermion flow~\cite{Denner:1992vza}  in Fig.~\ref{fig:diagram2}.
In a similar manner,  the Feynman amplitude of the RHN decay having the identical three body final states for which graviton emission occurs from Higgs side is given by, 
\begin{align}
	& \mathcal{M}_2=-\frac{Y_\nu r_\mu r_\nu}{M_p (r.k)}\left[\bar{u}_\ell(q)\mathbb{P}_R u_N(p)\right] \epsilon^{*\mu\nu}.
\end{align}	
Subsequently, using $q_2= k+q$ and summing over spins and polarization of fermions and gravitons respectively,  the  $\sum_{s, pol.}|\mathcal{M}_1|^2$ takes the form
\begin{align}
	\sum_{s,~ pol.}|\mathcal{M}_1|^2&=\frac{(Y_{\nu}^\dagger Y_{\nu})_{ii}}{4 M_p^2 (q.k)^2}\sum_{pol.} \epsilon^{\alpha \beta} \epsilon^{*\mu\nu} q_\mu q_\beta 
	\text{Tr}\left[(\slashed{q}+m_\ell)\gamma_\nu \mathbb{P}_L (\slashed{q_2}+m_\ell)\mathbb{P}_R(\slashed{p}+M_i)\mathbb{P}_L (\slashed{q_2}+m_\ell)\mathbb{P}_R \gamma_\alpha\right],
\end{align}
Similarly, for the $1 \rightarrow 3$ decay process of the RHN where graviton emission takes place from the Higgs side (right diagram of Fig.~\ref{fig:diagram2}), the squared Feynman amplitude are given by
\begin{align}
	\sum_{s,~pol.}|\mathcal{M}_2|^2&= \frac{(Y_{\nu}^\dagger Y_{\nu})_{ii}2(p.q)}{(r.k)^2 M_p^2}\sum_{pol.} \epsilon^{\alpha \beta} \epsilon^{*\mu\nu} r_\alpha r_\beta r_\mu r_\nu.
\end{align}
There should also exist interference term $\mathcal{M}_1\mathcal{M}_2^*$, which is estimated as
\begin{align}
	\sum_{s,~pol.}	\mathcal{M}_1 \mathcal{M}_2^*= \frac{(Y_\nu^\dagger Y_\nu)_{ii}}{2 (q.k)(r.k) M_p^2}
	\sum_{pol.}
	\text{Tr} \left[(\slashed{q}+m_\ell)\gamma_\nu \mathbb{P}_L (\slashed{q_2}+m_H)\mathbb{P}_R (\slashed{p}+M_i)\mathbb{P}_L\right] 
	q_\mu r_\alpha r_\beta \epsilon^{\alpha \beta}\epsilon^{*\mu\nu}.
\end{align}

The differential decay rate can then be evaluated as
\begin{align}
	\frac{d\Gamma^{1\to 3}}{d E_k}&=\frac{1}{8 M_i}\frac{1}{(2 \pi)^3}\int_{E_{q,min}}^{E_{q,max}} \sum_{pol.}\left(|\mathcal{M}_1|^2+|\mathcal{M}_2|^2+ \mathcal{M}_1 \mathcal{M}_2^*+ \mathcal{M}^*_1 \mathcal{M}_2\right) dE_q,
\end{align}
where the limits of the integration is given by
\begin{align}
	&E_{q,max/q,min}=\frac{M_i}{2(1-2 x)}\left[(1-3x +2 x^2-y_1^2+x y_1^2+y_2^2-x y_2^2)\pm x \alpha\right],\notag\\	
	&\alpha=\left(1-4 x+4 x^2-2 y_1^2+ 4x y_1^2+y_1^4-2 y_2^2+4 x y_2^2-2 y_1^2y_2^2+y_2^4\right)^{1/2},
	\label{eq:cross}
\end{align}
with  $x= E_k/M_i$, $y_1= m_H/M_i$ and $y_2= m_l/M_i$. 
Finally,  in the zero-mass limit of Higgs and leptons i.e., $y_1= y_2\rightarrow 0$, the differential decay rate of the  three-body decay process of RHNs takes the form
\begin{align}
 \frac{d\Gamma^{1\to3}}{d E_k}&= \frac{M_i^2 (Y_{\nu}^\dagger Y_{\nu})_{ii}  (1-x)^2(1-2x)}{128 M_p^2 \pi^3 x}, 
 \label{eq:decay}
\end{align}
as presented in Eq.~\eqref{decay-to-graviton}.

It is perhaps pertinent to point out here that being a Majorana fermion, the RHN can also decay to the antilepton and Higgs doublets. Consequently, one can imagine graviton production from the outgoing antilepton and Higgs leg, which has been generated from the decay of RHN. The relevant Feynman diagram would be similar to the Fig.\ref{fig:diagram2} except now, one would have antilepton and Higgs doublets in the final states. The main change in the relevant differential decay rate will come from the Feynman rule for the $N-\bar{\ell}_L H^\dagger$ vertex, which is given by $-i Y_\nu^\dagger \mathbb{P}_L$. In that case, it turns out that the sole contribution to differential decay rate follows from the GW production from Higgs leg while the GW contribution from anti lepton leg becomes identically zero in the massless limit. This follows from the typical Lorentz structure of the SM interaction involving the $SU(2)_L$ lepton doublet, leading to the respective amplitude being proportional to the mass of the anti lepton. The relevant differential decay rate for $N\to \bar{\ell}_L H^\dagger h$ can be expressed as
\begin{align}
	\frac{d\Gamma^{1 \to 3}}{d E_k}&= \frac{M_i^2 (Y_{\nu}^\dagger Y_{\nu})_{ii} (2-x)(1-2x)^2}{768 M_p^2 \pi^3 x},
\end{align}
which remains subdominant compared to Eq.~\eqref{eq:decay}. As a result, the graviton production will mostly be governed by three-body decay of RHN to lepton, Higgs and graviton, which matches the expression  given in \cite{Murayama:2025thw}.

\twocolumngrid

\bibliography{ref_1.bib}

\end{document}